# High resolution electron backscatter diffraction study on the heterogeneities of tetragonal distortion in Fe-C martensite at the microstructural scale

Tomohito Tanaka[a,b,*], Nozomu Nakamura[c], Angus J. Wilkinson[a]

[a] Department of Materials, University of Oxford, Parks Road, Oxford, UK
[b] Nippon Steel Corporation, 20-1 Shintomi, Futtsu, Chiba, Japan
[c] NS Solutions Corporation, 20-1 Shintomi, Futtsu, Chiba, Japan

* corresponding author.
E-mail: tanaka.m9p.tomohito@jp.nipponsteel.com

## Abstract

The spatial variation in martensite tetragonality ($c/a$ ratio) in Fe-0.77C (wt.%) alloy was investigated by means of pattern matching of electron backscatter diffraction patterns combined with high angular resolution electron backscatter diffraction analysis. It was found that the $c/a$ ratio varies within a martensite block and between blocks. The $c/a$ variation within a block is particularly evident in the vicinity of grain boundaries and shear strains are also present in addition to tetragonal distortion. The $c/a$ variation between blocks is more apparent than that within a block. The lattice parameter frequency profile predicted by our approach well matches with the X-ray diffraction profile from the same material by assuming reasonable residual strain acting on $a$- and $b$-axes. The heterogeneities of the crystal distortion are brought by a decrease and scatter in solid solution carbon and in carbon ordering as a result of the martensite transformation sequence as well as heterogeneous residual strain.

Graphical Abstract

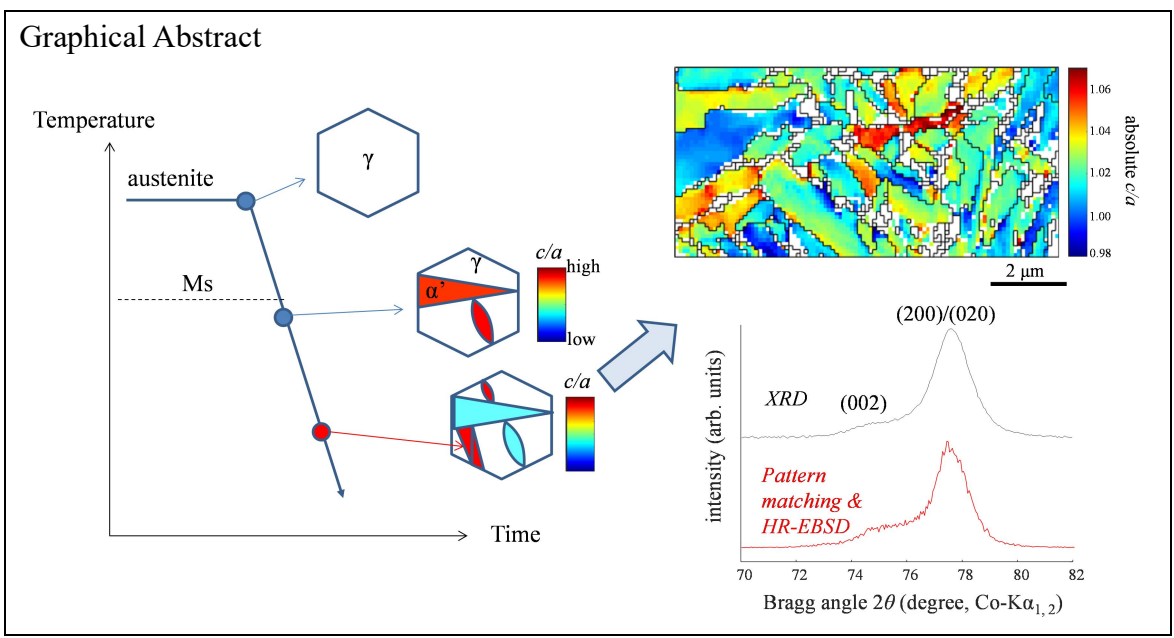

# 1. Introduction

The utilisation of quenched hard martensite with high volume fraction is becoming indispensable for the development of high strength steels [1]. However it is known that the hard martensite phase with medium to high carbon content shows less ductility and high susceptibility to hydrogen embrittlement [1-6]. Recently the type II residual strain introduced by martensitic transformation is drawing attention as one of the factors in determining the mechanical properties of Fe-C martensite such as initial yielding and uniform elongation [7-10]. Therefore the crystal structure and lattice parameter variation (*i.e.* strain) at microstructural length scale in Fe-C martensite is of great interest.

In order to investigate the local variation in the tetragonality (*c/a* ratio) of Fe-C martensite, several Fe-C martensite with different carbon contents were analysed by the authors with X-ray diffraction (XRD) and pattern matching of electron backscatter diffraction patterns (EBSPs) [11, 12]. It was found that the local *c/a* ratio varies by several per cent in room temperature (RT) aged high carbon steels with the averaged *c/a* ratio being linearly dependent on the nominal carbon content [12]. The averaged *c/a* ratio with respect to the carbon content was somewhat smaller than expected [12] from the classical model [13-16]. This analysis was carried out with a careful calibration of EBSD system using a strain-free calibrant mounted onto the martensite specimens [12]. The heterogeneity of local *c/a* ratio was analysed when the electron beam was focused on random locations on the martensite surface.

In order to map the spatial variation in *c/a* ratio in Fe-C martensite, an approach combining the pattern matching of EBSPs [11, 12] and high angular resolution electron backscatter diffraction (HR-EBSD) [17] is explored in this study. The application of pattern matching, or dictionary indexing [18, 19], of EBSPs for all analysis points would be enough to map the *c/a* ratio of martensite. However, HR-EBSD can offer more accurate and additional information on the distorted nature of crystal structures in Fe-C martensite rather than just tetragonal distortion. It is noted that local crystal structure of Fe-C martensite can be anticipated to have lower symmetry than simple body centred tetragonal (BCT) structure [12, 20, 21].

HR-EBSD enables the measurement of *relative* strain compared to the strain at a reference point which is often unknown [17] thus allowing for the determination of *relative c/a* distribution. Since the *c/a* ratio at reference points can be determined by the pattern matching of EBSPs [11, 12], the *absolute c/a* ratio can be deduced by simply offsetting the *relative c/a* ratio distribution. In this study, the *c/a* ratio mapping in Fe-0.77 wt.% C (hereafter Fe-0.77C) was carried out with this approach. The validity and limitation of the *c*-axis assessment through the pattern matching approach were also investigated. Furthermore the calculated *c/a* ratio distribution is related to microstructural features such as the shortest distance to the nearest grain boundaries (GBs) and geometrically necessary dislocation (GND) densities. Since carbon distribution heterogeneity in Fe-C martensite may be closely linked to the microstructural features (*e.g.* segregation of carbon to boundaries/dislocations [22], carbide precipitation according to the size of laths [23]), the spatial variation in the *c/a* ratio would give a deep insight into the behaviour of carbon during martensitic transformation. The redistribution behaviour of carbon atoms in martensite is discussed in the light of these new EBSD analysis results.

## 2. Methodology

2-1. Materials

We prepared an Fe-0.77C-1.00Mn (in wt.%) alloy and interstitial free (IF) steel. The IF steel was prepared to obtain an EBSD pattern from strain-free BCC ferrite for the calibration of the EBSD system [12]. Both steels were vacuum melted, hot rolled and cold rolled down to the thickness of approximately 1 *mm*. Then the Fe-0.77C alloy was austenitised at 1100 °C for 10 *min* and water-cooled to obtain martensite microstructure. Sub-zero treatment using liquid nitrogen then followed to reduce the amount of retained austenite. After that, the martensite specimens were kept at RT until EBSD measurements for ~3 months [12].

After the cold rolling process, the IF steel was annealed at 800 °C for 10 *min* for recrystallisation. Both Fe-0.77C martensite and IF steel specimens were cut and polished with SiC papers followed by colloidal silica (Bheuler, MasterMet) polishing for EBSD measurement.

The IF steel was annealed at 700 °C for 30 *min* in vacuum to remove potential residual strain introduced by the polishing. A thin foil with dimensions of 20 $\mu m$ × 20 $\mu m$ × 1 $\mu m$ (thickness: 1 $\mu m$) was extracted from the IF steel with a Ga focused ion beam (FIB) microsampling apparatus (Hitachi FB2000) [12]. The acceleration voltage of the Ga ion gun was 40 *kV*, followed by 10 *kV* to reduce the thickness of Ga ion beam damaged layer. The extracted thin foil specimen was glued on the martensite surface by depositing Pt layer at the edge of the foil. After that, in order to remove Ga-ion beam damaged layer near the foil surface, broad Ar ion beam with acceleration voltage of 1 *kV* was employed for 5 minutes [12].

2-2. (HR-)EBSD measurement

The EBSPs were collected using an SEM-EBSD system (SEM: JEOL 7100F, EBSD: EDAX DigiView camera). The sample was inclined 70° from horizontal and the phosphor screen was tilted 3° from the vertical. The acceleration voltage of the field emission gun was 20 *kV* with beam current of 14 *nA*. The EBSPs were recorded with 956 × 956 pixels resolution using the circular phosphor screen, and an integration time of about 0.5 s/pattern. Static background was subtracted from a raw EBSP. The orientation map for Fe-0.77C was obtained by scanning electron beam over 5 μm × 10 μm area with a step size of 0.1 μm. The pattern centre position was determined through pattern matching approach [11] using the experimental pattern from strain-free calibrant mounted on the martensite surface and the beam shift correction [12].

Crystal orientations are characterised with conventional Hough transform based analysis using *TSL OIM Data Collection* software (EDAX). Three kinds of assumptions on the crystal structure of Fe-C martensite were made; namely body centred cubic (BCC), body centred tetragonal (BCT) with single *c/a* ratio of 1.02, and BCT with multiple (17) possible *c/a* ratios from 0.99 to 1.10. At each analysis point, the *c/a* ratio with either the maximum votes or the minimum Fit value was selected within the BCT indexing with multiple possible *c/a* ratios. A Fit parameter defines the average angular deviation between the calculated bands and the detected bands. The lattice parameters used to create the lookup tables are listed in Table 1. The pixel size of EBSPs was reduced down to 96×96 pixels for the Hough transform analysis with a Hough resolution of 0.25°. Convolution mask size of 9×9 was used. The positions of 10 detected Kikuchi bands were used for the orientation calculation. The calculated crystal orientations were plotted on (inverse) pole figures using *TSL OIM Analysis* software (EDAX) or MATLAB *MTEX* toolbox [24].

Table 1. Crystal structure and lattice parameters used for the indexing with *TSL OIM Data Collection* software.

| indexing type | assumed crystal structure | lattice parameter (*nm*) | | *c/a* ratio |
|---|---|---|---|---|
| | | *a, b*-axes | *c*-axis | |
| BCC indexing | BCC | 0.2866 | 0.2866 | 1.00 |
| BCT indexing | BCT with single *c/a* | 0.2866 | 0.2926 | 1.02 |
| | BCT with multiple *c/a* | 0.2866 | 0.2826 ~ 0.3146 (0.020 *nm* step) | 0.99 ~ 1.10 |

Next, elastic strain analysis was performed using commercially available software *CrossCourt4* (BLG Vantage) for unbinned EBSPs. 30 regions of interests (ROIs) with 256 × 256 pixels size were selected over each EBSP image and cross correlation calculation was carried out in the Fourier domain after high/low pass filtering [17]. Reference EBSPs were selected by eye so as not to select overlapped patterns as references. The remapping technique [25] was applied to remove the undesired influence of larger lattice rotation on the elastic strain measurement. Mean angular error (MAE) and geometric mean of cross-correlation (XCF) peak heights [26] between a target and reference EBSPs were also assessed and the strain analysis results at pixel positions with MAE > 0.02 or the geometric mean of XCF < 0.7 were omitted from further analysis.

The elastic stiffness constants of Fe-C martensite are dependent on the carbon content [27]. It is not clear if the elastic stiffness constants show spatial variation in the Fe-C martensite. We used the elastic stiffness constants for BCC Fe reported in [28], namely $c_{11}$ = 231.5 GPa, $c_{12}$ = 135.0 GPa and $c_{44}$ = 116.0 GPa [28]. The elastic stiffness constants for Fe-0.5C martensite with BCT crystal structure, reported in [28] ($c_{11}$ = 268.1 GPa, $c_{12}$ = 111.2 GPa, $c_{13}$ = 110.2 GPa, $c_{33}$ = 267.2 GPa, $c_{44}$ = 79.06 GPa and $c_{66}$ = 78.85 GPa), were also tested to check the elastic strain error caused by the choice of elastic stiffness constants. The error was found to be on the order of $10^{-4}$ (see Supplementary materials) and this level of error did not cause a major problem with the *c/a* ratio measurement results shown in Results and Discussion section.

The density of GNDs in Fe is calculated from the lattice rotation components obtained by the above HR-EBSD cross-correlation analysis, assuming that the crystal structure of Fe is BCC. The dislocations are assumed to have *a*/2 <111> Burgers vector slipping on {110} planes. Four screw types with line direction <111> and twelve edge types with line direction <112> are considered [29].

2-3. Procedures for absolute tetragonality distribution assessment

*Absolute* tetragonality at an analysis point, $(c/a)_{abs}$, in an EBSD map is calculated using the following equation [30].

$$\left(\frac{c}{a}\right)_{\text{abs}} = \left(\frac{c}{a}\right)_{\text{ref}} + e_{33}^{\text{crystal}} - \frac{e_{11}^{\text{crystal}} + e_{22}^{\text{crystal}}}{2} \tag{1}$$

where $(c/a)_{\text{ref}}$ is *absolute* tetragonality at the reference point for the grain to which the analysis point belongs. $e_{ij}^{\text{crystal}}$ (i, j = 1, 2, 3) is *relative* elastic strain tensor on crystal frame compared to the strain at the reference point. $e_{11}^{\text{crystal}}, e_{22}^{\text{crystal}}, e_{33}^{\text{crystal}}$ are *relative* elastic strain acting on *a*-, *b*-, *c*-axis, respectively, at the analysis point. $(c/a)_{\text{ref}}$ is determined through the pattern matching of EBSPs described fully in the previous paper [12]. $e_{ij}^{\text{crystal}}$ is calculated from the coordinate transformation of the *relative* elastic strain tensor on sample frame, $e_{ij}^{\text{sample}}$, using the rotation matrix $\boldsymbol{R}_{\text{s}\to\text{c}}^{\text{TSL}}$ and its transpose $\left(\boldsymbol{R}_{\text{s}\to\text{c}}^{\text{TSL}}\right)^t$. $\boldsymbol{R}_{\text{s}\to\text{c}}^{\text{TSL}}$ describes the rotation from the sample frame (Fig. 1(a)) into the crystal frame.

$$e_{ij}^{\text{crystal}} = \boldsymbol{R}_{\text{s}\to\text{c}}^{\text{TSL}} \, e_{ij}^{\text{sample}} \left(\boldsymbol{R}_{\text{s}\to\text{c}}^{\text{TSL}}\right)^t \tag{2}$$

$e_{ij}^{\text{sample}}$ is obtained using the *CrossCourt4* software (BLG Vantage) implementation of HR-EBSD mapping[1]. The rotation matrix $\boldsymbol{R}_{\text{s}\to\text{c}}^{\text{TSL}}$ is obtained for each point in the map using a set of given Euler angles ($\varphi_1$, $\Phi$, $\varphi_2$) determined by the Hough transform based analysis with *TSL OIM Data Collection* software as follows:

$$\boldsymbol{R}_{\text{s}\to\text{c}}^{\text{TSL}} = \begin{pmatrix} \cos\varphi_2 & \sin\varphi_2 & 0 \\ -\sin\varphi_2 & \cos\varphi_2 & 0 \\ 0 & 0 & 1 \end{pmatrix} \begin{pmatrix} 1 & 0 & 0 \\ 0 & \cos\Phi & \sin\Phi \\ 0 & -\sin\Phi & \cos\Phi \end{pmatrix} \begin{pmatrix} \cos\varphi_1 & \sin\varphi_1 & 0 \\ -\sin\varphi_1 & \cos\varphi_1 & 0 \\ 0 & 0 & 1 \end{pmatrix} \tag{3}$$

Since the Hough transform based analysis may fail to distinguish *c*-axis from *a*, *b*-axes due to the poor angular resolution, the rotation matrix $\boldsymbol{R}_{\text{s}\to\text{c}}^{\text{TSL}}$ may need to be corrected by substituting the rotation matrix determined by the pattern matching of EBSPs, $\boldsymbol{R}_{\text{r}\to\text{c}}^{\text{PM}}$, for $\boldsymbol{R}_{\text{s}\to\text{c}}^{\text{TSL}}$. This is because the pattern matching approach delivers more accurate results on *c*-axis assessment, which will be shown later. In this study $\boldsymbol{R}_{\text{r}\to\text{c}}^{\text{PM}}$ was calculated at one analysis point (*i.e.* reference point) per one grain and $\boldsymbol{R}_{\text{s}\to\text{c}}^{\text{TSL}}$ for all the analysis points in the grain was corrected so that the *c*-axes determined by the TSL software and by the pattern matching approach become close to identical. In order to do this, one of the following matrices needs to be multiplied by $R_{s \to c}^{TSL}$.

---

[1] It is noted that the orthogonal reference frames used in the *CrossCourt4* software and in TSL software are different. In this study, the strain tensor obtained in the *CrossCourt4* was coordinate transformed before Eq.(2) such that it is described on the sample frame (Fig. 1(a)).

$$S_{axis(l)} = \begin{pmatrix} 1 & 0 & 0 \\ 0 & 1 & 0 \\ 0 & 0 & 1 \end{pmatrix}, \begin{pmatrix} 0 & 0 & 1 \\ 0 & 1 & 0 \\ -1 & 0 & 0 \end{pmatrix}, \begin{pmatrix} 1 & 0 & 0 \\ 0 & 0 & 1 \\ 0 & -1 & 0 \end{pmatrix}, l \in \{1,2,3\}. \tag{4}$$

where the first matrix of $S_{axis}$ is identity matrix, the second matrix swaps *a*-axis with *c*-axis and the third matrix swaps *b*-axis with *c*-axis while maintaining the right-handed coordinate. Then the following equation holds if and only if $S_{axis(l)}$ and a symmetry operator matrix for BCT, $S_{tet(k)}$ (where $k \in \{1,2,\ldots,16\}$), are properly chosen.

$$S_{tet(k)} S_{axis(l)} R_{s \to c}^{TSL} = R_{r \to c}^{PM} g \tag{5}$$

where $g$ is the 3×3 orthogonal matrix which describes the rotation from the sample frame used in the *TSL OIM Data Collection* software into the reference frame used in the pattern matching of EBSP analysis. The geometrical configuration between the two orthogonal frames is shown in Fig. 1 and then $g$ is as follows:

$$g = \begin{pmatrix} 0 & 1 & 0 \\ -\cos\tau & 0 & \sin\tau \\ \sin\tau & 0 & \cos\tau \end{pmatrix} \tag{6}$$

where $\tau = 90 - \tau_{sample} + \tau_{detector}$. $\tau_{sample}$ is the sample tilt, 70 degrees, and $\tau_{detector}$ is the detector tilt, 3 degrees in this study. Symmetry operator matrices for BCT crystal structure, $S_{tet(k)}$, are listed below.

$$S_{tet(k)} = \begin{pmatrix} \pm 1 & 0 & 0 \\ 0 & \pm 1 & 0 \\ 0 & 0 & \pm 1 \end{pmatrix}, \begin{pmatrix} 0 & \pm 1 & 0 \\ \pm 1 & 0 & 0 \\ 0 & 0 & \pm 1 \end{pmatrix}, k \in \{1,2,\ldots,16\}. \tag{7}$$

However, in reality, no unique set of $S_{axis(l)}$ and $S_{tet(k)}$ can satisfy the Eq. (5) because of inevitable measurement errors. Thus the subscripts $k$ and $l$ are determined by the minimisation of Frobenius norm of the difference between the left- and right-hand sides of the Eq. (5).

$$\{k\},\{l\} = \underset{k\in\{1,2,...,16\},\, l\in\{1,2,3\}}{\arg\min} \left\{ \left\| \boldsymbol{S}_{\text{tet}(k)}\, \boldsymbol{S}_{\text{axis}(l)}\, \boldsymbol{R}_{\text{s}\to\text{c}}^{\text{TSL}} - \boldsymbol{R}_{\text{r}\to\text{c}}^{\text{PM}}\, \boldsymbol{g} \right\| \right\} \tag{8}$$

Once $k$ and $l$ are determined, the rotation matrix $\boldsymbol{R}_{\text{s}\to\text{c}}^{\text{TSL}}$ is replaced with $\boldsymbol{S}_{\text{tet}(k)}\, \boldsymbol{S}_{\text{axis}(l)}\, \boldsymbol{R}_{\text{s}\to\text{c}}^{\text{TSL}}$ to calculate $e_{ij}^{\text{crystal}}$ using Eq. (2), followed by the calculation of $(c/a)_{\text{abs}}$ using Eq. (1). These procedures are summarised in the flowchart presented in Fig. 2.

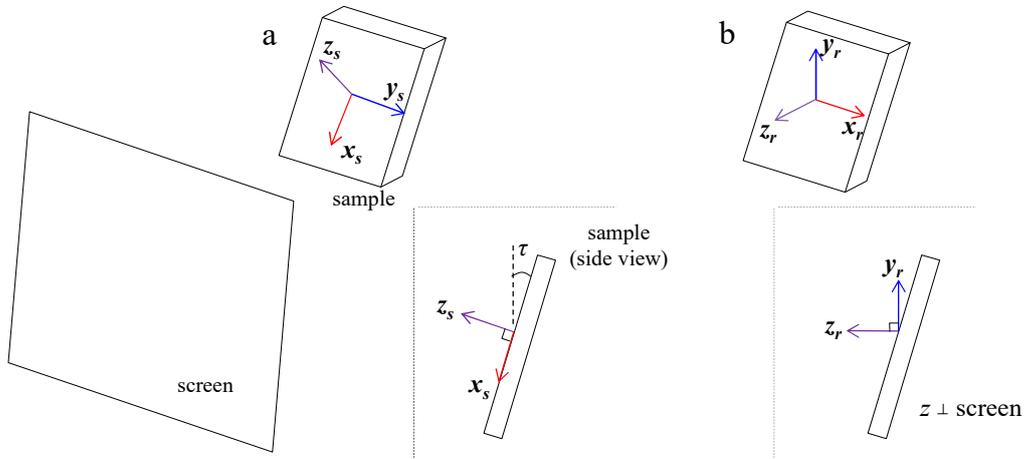

**Figure 1**. (a) Sample frame used in the *TSL OIM Data Collection* software, (b) Reference frame used within our pattern matching approach [11].

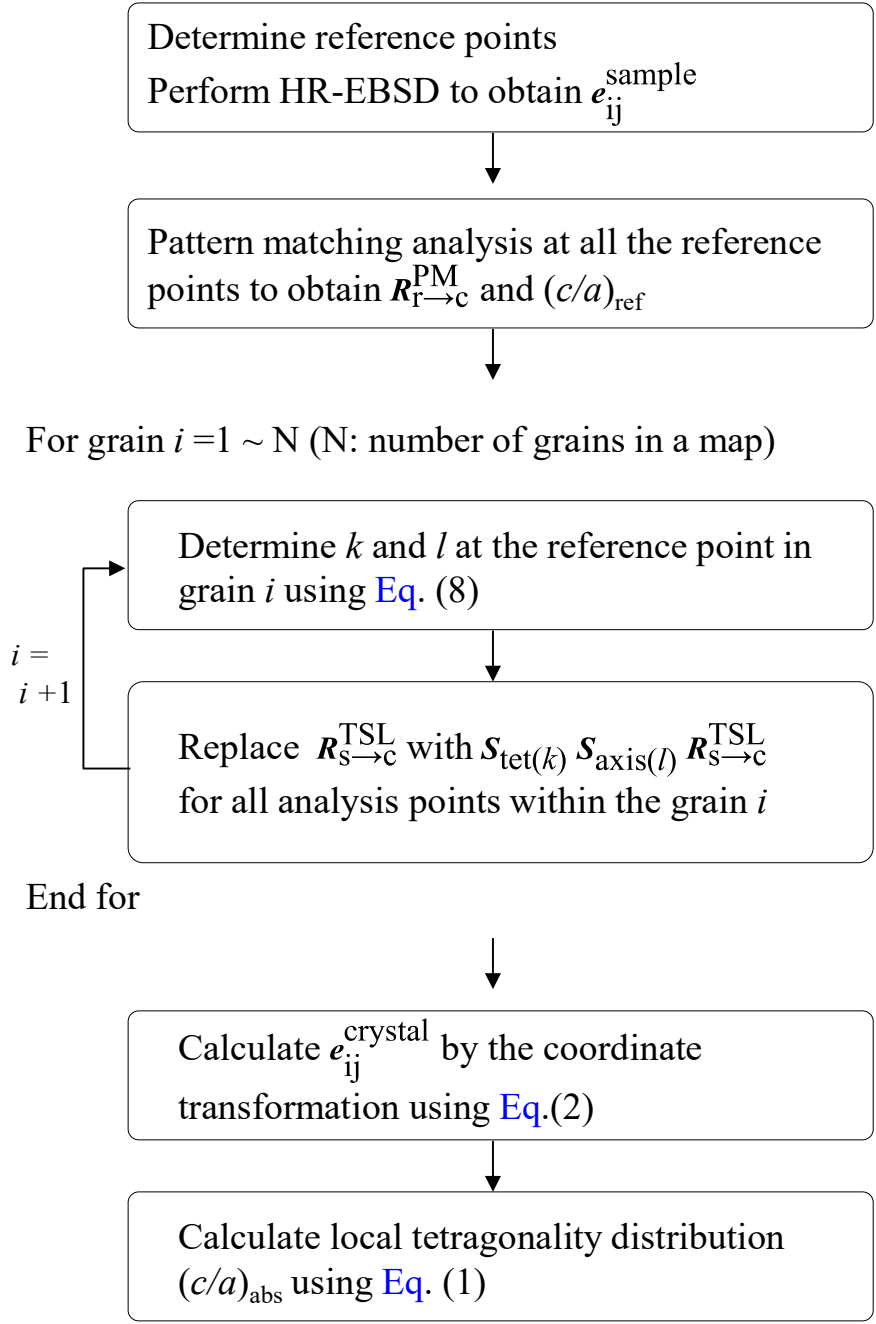

**Figure 2**. Flowchart for the absolute *c/a* ratio calculation.

## 3. Results and Discussion

3-1. Problem with conventional Hough transform based indexing

Figs. 3(a)-3(b) shows the normal direction inverse pole figure (IPF-ND) map and the Bain group map for quenched and RT-aged Fe-0.77C martensite obtained with the BCC indexing. The colours used in Fig. 3(b) are based on the colour assignments for the three Bain circles in the {001} pole figure shown in Fig. 3(c). The EBSD data are found to be obtained from a single austenite grain as only three Bain circles are seen in Fig. 3(c). The centre of each circle corresponds to {100} family plane in the prior austenite according to the Bain correspondence [31] ($[1\bar{1}0]_\gamma$ // $[100]_{\alpha\text{-BCT}}$, $[110]_\gamma$ // $[010]_{\alpha\text{-BCT}}$, $[001]_\gamma$ // $[001]_{\alpha\text{-BCT}}$). Therefore the grains with the same colour in Fig. 3(b) should have the common BCT *c*-axis. However, the differentiation of the *c*-axis from the *a, b*-axes is not possible as the crystal structure is assumed to be BCC.

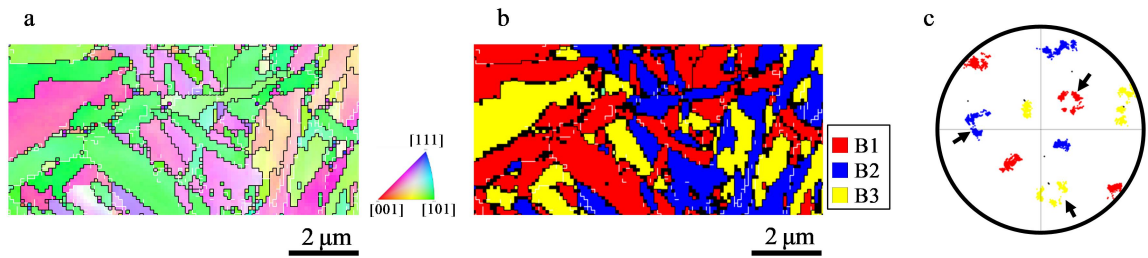

**Figure 3**. (a) IFP-ND for Fe-0.77C martensite. The GBs with misorientation angle > 15º and 2º ~ 15º are shown as black lines and white ones respectively. (b) Bain group classification. (c) {001} pole figure for (a). Three arrows indicate the location of the Bain circles. The data points with confidence index (CI) value < 0.2 were omitted.

Figs. 4(a), (d), and (g) shows the IPF-ND map obtained with BCT indexing for the same map of EBSPs used to construct Fig. 3(a). With BCT indexing, a patchy colour distribution within a grain determined by the BCC indexing is obtained because of the uncertainty in the determination of *c*-axis. The abrupt change in the colour within a grain corresponds to 90 degrees rotation. The patchy pattern is more evident with the BCT indexing with varying *c/a* in Figs. 4(d) and (g). This indexing type allows for the *c/a* mapping, showing the heterogeneous distribution of *c/a* in Figs. 4(e) and (h). It is unlikely that the patchy *c/a* distribution in Fig. 4(e) reflects true distribution because of the discontinuities in *c/a* between neighbouring pixels in a grain. The *c/a* distribution determined by the minimum Fit approach (Fig. 4(h)) shows that the *c/a* ratio seems to be different between grains rather than between grain interiors.

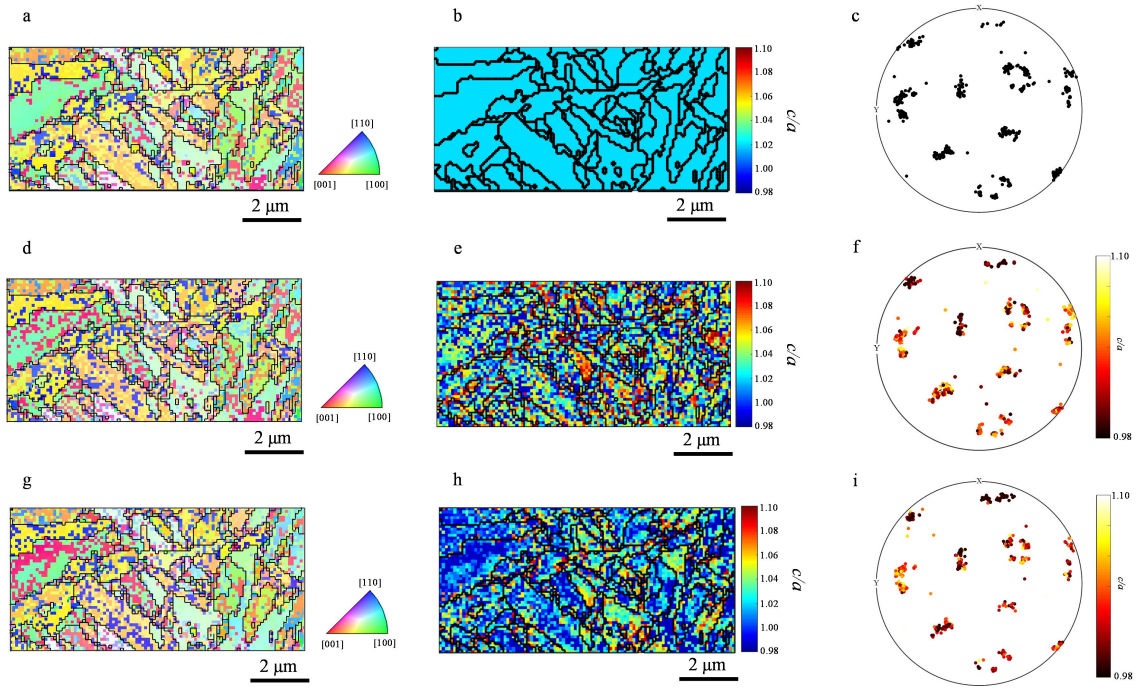

**Figure 4**. (a), (d), (g) IPF-ND map, (b), (e), (h) *c/a* map, and (c), (f), (i) (001) pole figure for quenched and RT aged Fe-0.77C martensite analysed by the Hough transform based analysis. (a)~(c) BCT indexing with fixed *c/a* of 1.02, (d)~(i) BCT indexing with varying *c/a* from *c/a* = 0.99 to *c/a* = 1.10. The local *c/a* ratio was determined with (d)~(f) maximum votes and with (g)~(i) minimum Fit value. (c), (f), (i) The same data points in Fig. 3(c) with CI value < 0.2 obtained by the BCC indexing were removed.

Figs. 4(c), (f), and (i) depict the (001) pole figures for the data points in Figs. 4(a), (d) and (g) respectively. If the BCT indexing works perfectly, all the data points in the (001) pole figure should be on one of the three Bain circles. However, a large fraction of data points are plotted off the Bain circle. In order to evaluate the reliability of the *c*-axis assessment with the Hough transform based analysis, the success rate of BCT indexing are calculated based on the Bain correspondence. The success rate is defined [32] as a ratio of the number of data points, whose [001]*α*-BCT is contained in the Bain circle, to the total number of data points. Table 2 shows the influence of indexing type on the success rate and the averaged Fit parameter. It is found that the BCT indexing with single *c/a* methods show the success rate of nearly 50 %, which agrees well with the value reported in [32]. Consideration of varying *c/a* does not improve the success rate and the averaged Fit due to the lack in angular resolution of the Hough transform based indexing. Therefore, the reliable differentiation of *c*-axis from the *a, b*-axes is found to be difficult with conventional Hough transform based analysis when the carbon content is 0.77 wt.%.

**Table 2**. Success rate of *c*-axis assessment according to the indexing type using Hough transform based analysis.

| indexing type | success rate of *c*-axis assessment | Fit (degree) | average *c/a* |
|---|---|---|---|
| BCC indexing | (33 %) | 1.3 ± 0.4 | 1 |
| BCT indexing with single *c/a* | 51 % | 1.5 ± 0.4 | 1.02 |
| BCT indexing with multiple *c/a* (maximum vote) | 44 % | 1.5 ± 0.4 | 1.03 |
| BCT indexing with multiple *c/a* (minimum Fit value) | 45 % | 1.4 ± 0.3 | 1.03 |

3-2. Validity and limitation of *c*-axis assessment with pattern matching of EBSPs

The BCT indexing success rate with the pattern matching of EBSP is also evaluated. The experimental EBSPs were taken from 54 analysis points marked as the black squares in Fig. 5(a). These points are the reference points for HR-EBSD strain mapping. For comparison, the BCC indexing and BCT indexing with a single *c/a* of 1.02 using the conventional Hough transform were also employed for the same set of EBSPs. The (001) pole figures with BCC indexing, BCT indexing with fixed *c/a* ratio of 1.02 and BCT indexing with pattern matching approach are shown in Figs. 5(b), (c) and (d), respectively.

The success rates are listed in Table 3. It is found that the success rate of BCT indexing with the pattern matching approach is greatly increased compared to the Hough transform based indexing. This demonstrates an improvement in the angular resolution by using the pattern matching approach [11]. However, the success rate of 100 % is not achieved as can be seen in Fig. 5(d) showing some of the data points are still away from the Bain circles on the (001) pole figure.

Out of 54 data points on the (001) pole figure in Fig. 5(d), 14 data points are off the Bain circles. It is found that out of the 14 reference points, the *c/a* ratio at 11 reference points is smaller than or close to 1.00~1.01. When the *c/a* is close to unity, the *c*-axis is hardly distinguished from the *a*, *b*-axes as the length of all three principal axes are almost identical. Furthermore, the longest principal axis might be perturbed within a grain [20] when *c/a* is close to unity because of the presence of residual strain and carbon distribution heterogeneity. The *c/a* ratio at the other 3 reference points were calculated as more than 1.02. The reason for being off the Bain circle at these 3 points is not clear. It might be related to extremely large residual strain or to the wrong assumption made for the pattern matching calculation when the local crystal structure can be regarded as orthorhombic or lower symmetry structure. In any case it can be said that the main reason for being off the Bain circles at some reference points is the similarities of the lengths of *a, b* and *c*-axes at these analysis points.

This means that the BCT *c*-axis can be defined in two different ways. The *c*-axis defined by the Bain correspondence and the one determined by the pattern matching approach. The BCT *c*-axis defined by the Bain correspondence, (BCT *c*-axis)$_{Bain}$, is identical to the <001> *c*-axis of the parent austenite phase with face centred cubic (FCC) crystal structure[2] while the *c*-axis determined by the pattern matching EBSD analysis, (BCT *c*-axis)$_{EBSD}$, defines it as the longest axis among all three principal axes. (BCT *c*-axis)$_{Bain}$ and (BCT *c*-axis)$_{EBSD}$ should be identical when the martensite is 'fresh', but because of the carbon redistribution during cooling (i.e. auto-tempering) and RT aging, the length of *c*-axis can be shortened [34]. The compressive residual strain acting on the (BCT *c*-axis)$_{Bain}$ could also contribute to a decrease in the *c*-axis length [35-37]. Similarly, the length of *a*, *b*-axes could be elongated because of the large tensile residual strain acting on the *a*, *b*-axes or carbon occupation in interstitial octahedral *x/y* sites. Such conditions could cause the cases where (BCT *c*-axis)$_{Bain}$ and (BCT *c*-axis)$_{EBSD}$ are no longer identical. This may be more likely to happen in low carbon martensite as the length of the *c*-axis is shorter in the martensite with smaller carbon content and the extent of auto-tempering is more apparent for lower carbon martensite due to the higher *Ms* temperature.

The longest axis flip within a grain is examined in Fig. 6. Fig. 6(b) shows the pattern matching analysis results between dynamically simulated EBSPs [38] and the experimental EBSP from Point A in the green coloured grain (Fig. 6(a)). The dynamically simulated patterns for BCT Fe with varying $c/a$ are obtained considering the possibility of $c$-axis being along each of the three principal axes (axis 1, 2, 3) [12]. The negative of normalised cross correlation coefficient, $-r$, between simulated and experimental EBSPs is assessed as a similarity metric. The best pattern match solution (ie the lowest minima in $-r$) for Point A is found (Fig. 6(b)) for a $c/a$ ratio of ~1.03, with the longest axis, ie the (BCT $c$-axis)$_{EBSD,}$ corresponding to (axis 2) which lies on the Bain circle. Fig. 6(c) shows another pattern matching analysis results for Point B in the same grain. The same axis (axis 2) is also determined as the (BCT $c$-axis)$_{EBSD}$ , while the $c/a$ is somewhat higher at ~1.05, at this point. It appear that the (BCT $c$-axis)$_{EBSD}$ does not change, and is consistent with the Bain correspondence analysis in this grain with relatively large $c/a$. However, similar analysis for the blue coloured grain in Fig. 6(a) which has relatively low tetragonality, shows that the alignment of the longest (BCT $c$-axis)$_{EBSD}$ now changes from axis 1 at Point C to axis 2 at Point D. At both points the best match $c/a$ ratio is below 1.01, and while the long axis alignment lies on the Bain circle for point C it does not for point D. As mentioned above, in addition to carbon distribution heterogeneity, the elastic strain distribution is considered to contribute to the flip in the longest axis within a grain. The strain variation within a grain was analysed more completely with HR-EBSD measurement and is discussed in the following section.

---

[2] Actual orientation relationship (OR) between martensite and the parent austenite is not Bain correspondence and is known to deviate slightly from Kurjumov-Sacks OR [*e.g.* 33].

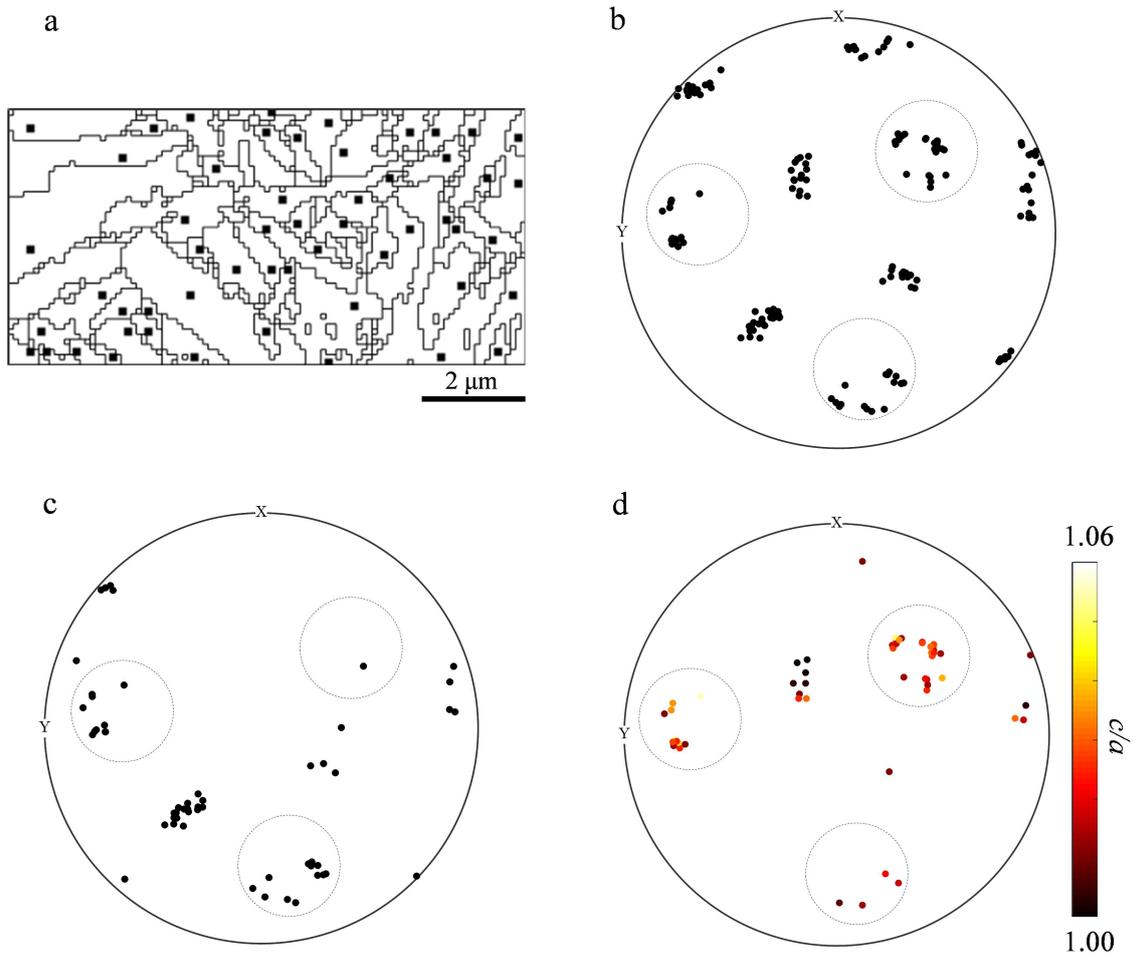

**Figure 5**. (a) The location of reference points marked as black squares (■). One reference point per one grain is selected. (b)-(d) (001) pole figures for Fe-0.77C martensite, obtained with (b) Hough transform based BCC indexing, (c) Hough transform based BCT indexing with fixed *c/a* of 1.02 and (d) pattern matching BCT indexing. The data points inside the dashed circles are on the Bain circles.

**Table 3.** Success rate of *c*-axis assessment for the reference EBSPs by the indexing with the Hough transform based analysis and with the pattern matching of EBSPs.

|  | indexing type | success rate of *c*-axis assessment | averaged *c/a* ratio | ±1 standard deviation |
|---|---|---|---|---|
| Hough transform based analysis | BCC indexing | 33 % | 1 | - |
|  | BCT indexing with single *c/a* | 41 % | 1.02 | - |
| pattern matching analysis | BCT indexing with multiple *c/a* | 74 % | 1.021 | 0.013 |

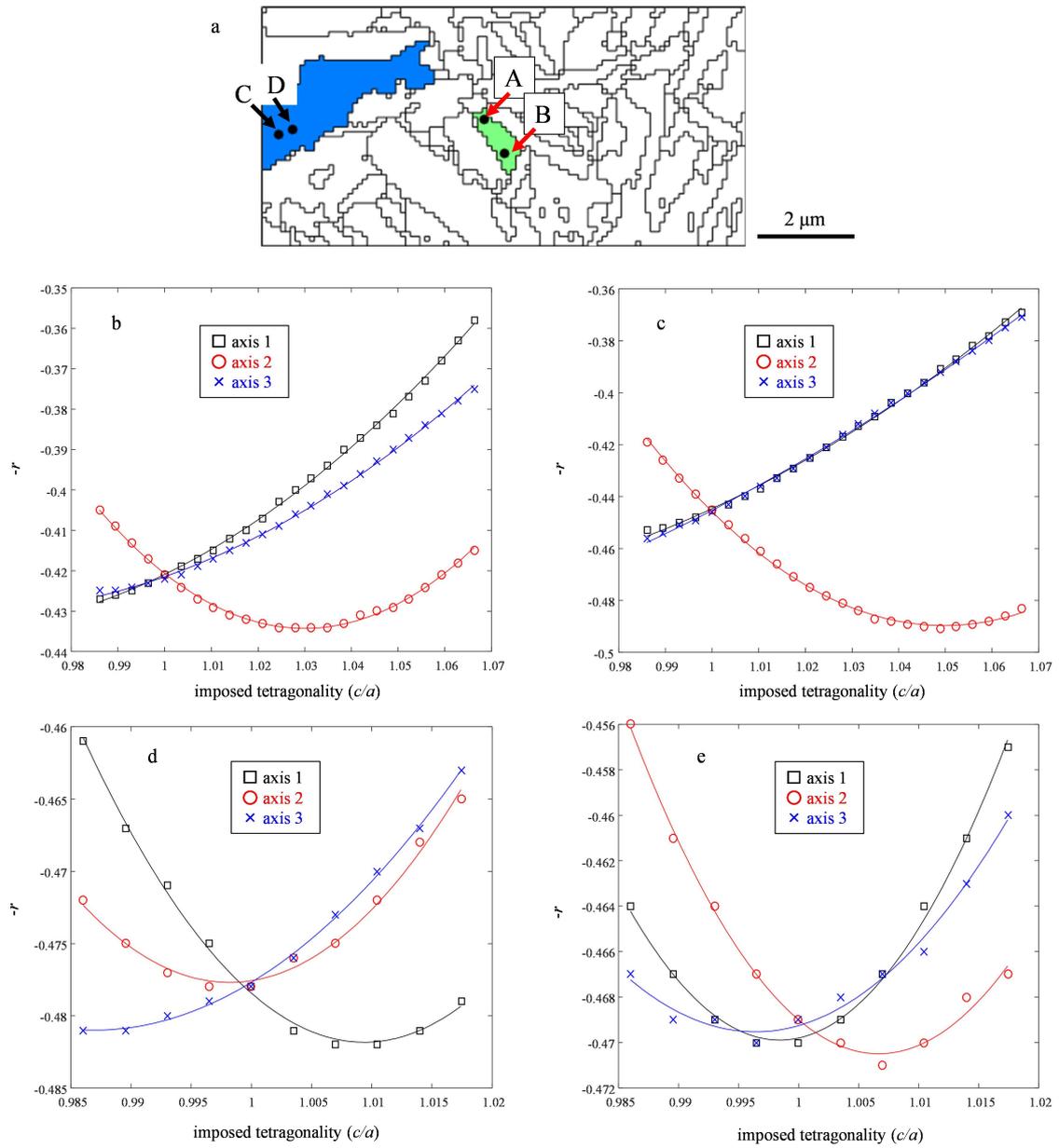

Figure 6. (a) Tetragonality analysis points using pattern matching of EBSPs, indicated by the arrows. (b)-(e) Tetragonality analysis results for (b) Point A, (c) Point B, (d) Point C and (e) Point D. Solid lines in the figures are the second order polynomial fitting results.

## 3-3. Tetragonality map analysed by the combination of pattern matching approach and HR-EBSD and its relation to microstructural features

Fig. 7 shows the relative strain map on crystal frame determined using Eq. (2) from the HR-EBSD data. The positions of reference points are identical to the ones in Fig. 5(a) used for the pattern matching analysis. The relative strain map shows the heterogeneous distribution of crystal distortion within a grain. In addition to the normal components, the shear strains are observed, indicating that the local crystal structure deviates from pure BCT. The strain deviation from grain-averaged strain is listed in Table 4. The ±1 standard deviation of ~ 0.4 - 0.6 % is significant. When the local value of ($c/a$ - 1) at reference points is similar in size to this standard deviation, the determined longest axis from the pattern matching approach can be flipped for some points within a grain. This agrees well with the results in Figs. 6(d) and (e). It can also be said that when the local ($c/a$ - 1) at reference points is larger than the deviation, the longest axis is unlikely to be flipped within the grain (see Figs. 6(b) and (c)).

This heterogeneity comes from both the local elastic strain and carbon distribution (solid solution carbon content and carbon ordering) heterogeneities, although the contribution of each factor to the crystal distortion is difficult to separate. Then the following two measures were assessed, namely $|e_{11}^{crystal} - e_{22}^{crystal}|$ as a measure of the extent of orthorhombicity and $|e_{12}^{crystal}| + |e_{23}^{crystal}| + |e_{31}^{crystal}|$ as a measure of the extent of shear distortion. The former is driven by differences in the occupancy of octahedral interstices $x$- and $y$-sites and by mechanical elastic strains while the latter comes from mechanically driven accommodation rather than being driven by carbon ordering. Figs. 8(a)-(c) show the maps and histograms of these two measures relative to the grain averaged values. The histogram (Fig. 8(c)) has a positively skewed distribution with the right tail being longer than the left because of the higher values localised in the vicinity of GBs. This was confirmed statistically by constructing a map of the Euclidian in plane distance to the nearest GB (Fig. 8(d)) and correlating this pixel by pixel with the two measures (Figs. 8(e) and (f)). These two measures apparently increases in the vicinity of GBs. Chen *et al* speculated in the previous literature [20] that carbon ordering of the octahedral sites might be perturbed at plate intersections where the local residual stresses vary. An increase in the extent of orthorhombicity near GBs would reflect this. The extent of shear distortion is also increased around GBs probably because of the constraint of shear deformation near GBs upon martensitic transformation. Therefore the deviation of the crystal structure from pure BCT is more evident near GBs because of the local mechanical stresses, and the influence of local carbon ordering heterogeneities on the deviation might be added.

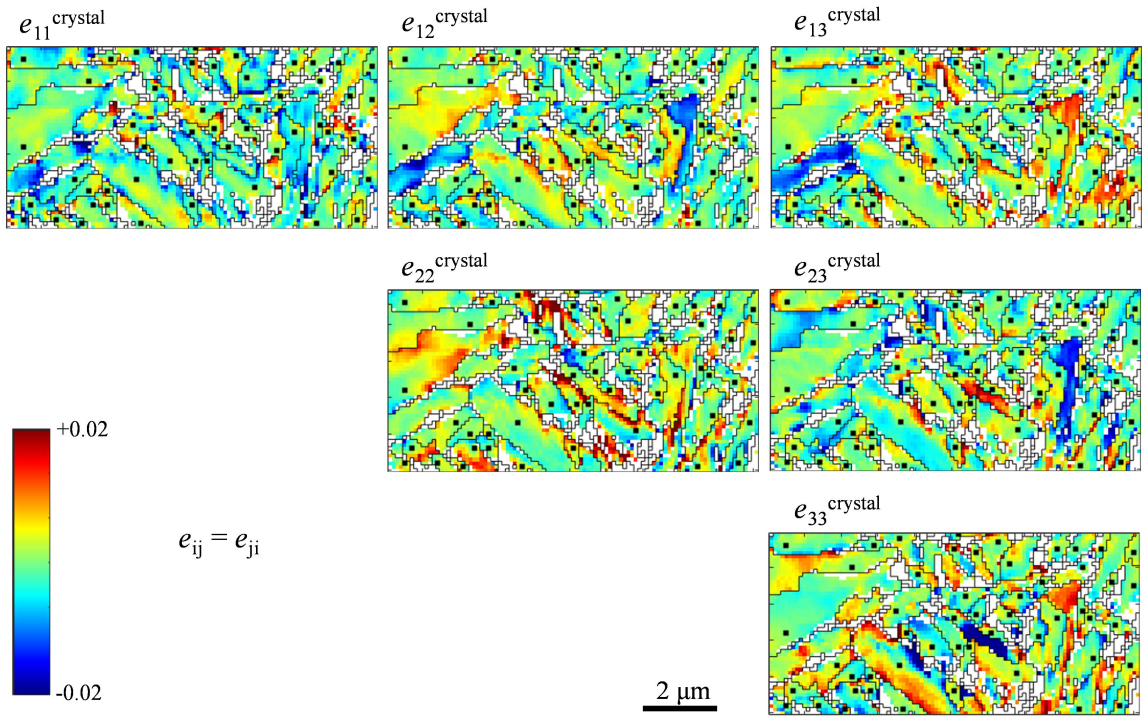

**Figure 7**. Relative strain map on crystal frame. The location of reference points are marked as '■'.

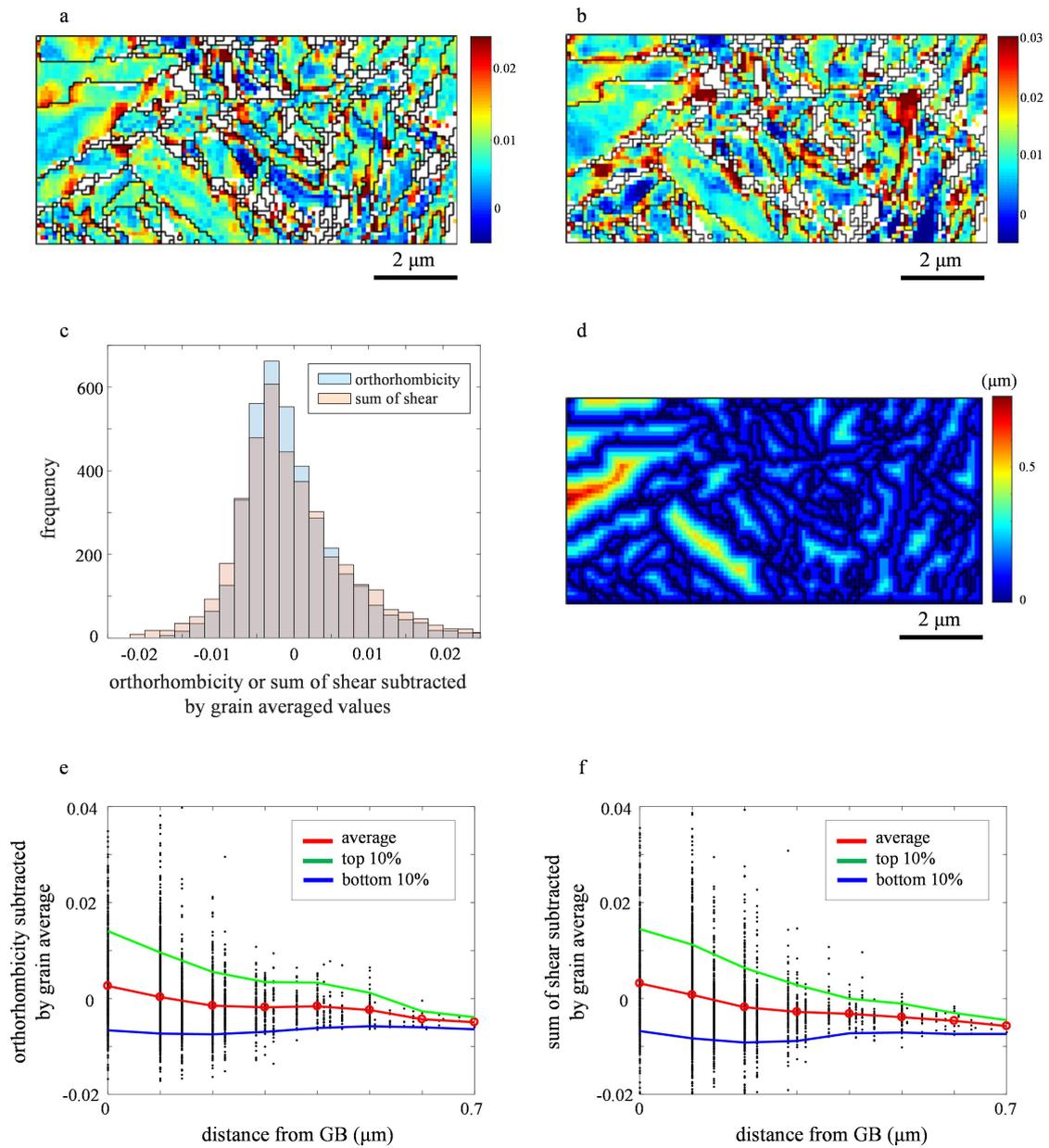

**Figure 8**. Maps of (a) orthorhombicity and (b) sum of shear subtracted by the grain averaged values. (c) Histograms of (a) and (b). (d) Shortest distance between each pixel and the nearest GB. (e) Orthorhombicity and (b) sum of shear subtracted by grain average, with respect to the distance to the nearest GB.

**Table 4.** The average and standard deviation of relative strain from grain averaged strain measured on crystal frame.

|        | $\Delta e_{11}$ | $\Delta e_{22}$ | $\Delta e_{33}$ | $\Delta e_{12}$ | $\Delta e_{31}$ | $\Delta e_{23}$ |
|--------|--------|--------|--------|--------|--------|--------|
| Average | 0.0000 | 0.0000 | 0.0000 | 0.0000 | 0.0000 | 0.0000 |
| ±1 std  | 0.0055 | 0.0061 | 0.0065 | 0.0045 | 0.0049 | 0.0055 |

The relative and absolute *c/a* map obtained using Eq.(1) are shown in Fig. 9, indicating that the *c/a* varies substantially within a grain (relative *c/a* = $e_{33}^{\text{crystal}} - \frac{e_{11}^{\text{crystal}} + e_{22}^{\text{crystal}}}{2}$ [30]). Table 5 summarises the average and standard deviation of analysed relative and absolute *c/a*. The average and standard deviation of absolute *c/a* for all analysed points in the map are nearly the same as that for just the reference points (see Table 3). However, the standard deviation of the relative *c/a* is smaller than that of the absolute *c/a*, indicating that the *c/a* variation between grains (*i.e.* blocks) is more marked than within a grain. In Fig. 9(b) the *c/a* ratios at some grains (red-coloured grains) exceed the one expected by the classical model (1 + 0.045 [C in wt.%] ~ 1.035). This is probably because the tensile elastic strain acting on the *c*-axis or compressive strain acting on the *a*- and *b*-axes contributes to an increase in the *c/a* ratio in the grains along with the carbon ordering. On the contrary, some other grains (dark blue-coloured grains) show the *c/a* ratio of nearly unity.

To validate our analysis results, the lattice parameter frequency profile obtained with the combination of pattern matching and HR-EBSD is compared to the XRD profile from the same material with nearly the same RT aging time. Now that the *c/a* ratio map is obtained, the lattice parameter frequency can be calculated by assuming the lattice parameters of *a*- and *b*-axes at the reference points. It is reasonable to assume that the the *a*- and *b*-axes lengths at all the reference points have an average of 0.2856 *nm*, which is determined by the XRD peak position [12], with standard deviation σ of certain values because of the presence of type II&III residual strains. Therefore normal random numbers with the average *m* and standard deviation *σ* were generated to assume the lattice parameters of *a*- and *b*-axes. Then the standard deviation of lattice strain *ε* is described as *σ/m*. The calculated lattice parameters were converted to the corresponding Bragg angle 2*θ* on the assumption that the target material for X-ray source is cobalt. This assumes a set of independently scattering domains in the XRD analysis volume and the effects of spatial variation in the strain distribution on the X-ray profile was not considered. Fig. 10 shows that the intensity profile with respect to the Bragg angle estimated by our EBSD method is in good accordance with the XRD profile when $\varepsilon = 5\times10^{-3}$. This standard deviation of $5\times10^{-3}$ well matches the values in Table 4 and the standard deviation of internal strains in as-quenched Fe-C martensite reported by Hutchinson *et al* [8]. Therefore it is found that the elastic strains in Fe-C martensite are surely necessary to reproduce the XRD profile. Furthermore our EBSD approach with careful calibration of the EBSD system [12] demonstrates the ability to measuring reasonable axial ratios at the microstructural length-scale, which is not possible by laboratory XRD apparatus.

The standard deviation of lattice strain acting on the *a*- and *b*-axes length, $5\times10^{-3}$, is nearly one third of that of the *c/a* ratio (Table 5) meaning that the deviation of *c*-axis length is more marked than *a*- and *b*-axes. Thus the variation of *c*-axis cannot be explained solely by the heterogeneous residual strain. Carbon distribution (carbon concentration and ordering) heterogeneities should play some role here.

The GND density was also determined from the HR-EBSD map and statistical values from the map are given in Table 5, while Figs. 11(a) and (b) shows the map and histogram of density values. The average of GND density in Fe-0.77C quenched martensite was estimated as $2.1\times10^{15}\ m^{-2}$, which is similar to the total dislocation density in Fe-0.78C quenched martensite ($2.4\times10^{15}\ m^{-2}$) reported from transmission electron microscope observations by Morito *et al* [39]. The GND density distribution is found to have significant spatial heterogeneities (Fig. 11(a)), similar to those reported in deformed polycrystalline metals that arise because of discontinuities in the slip system across GBs [40]. Like in deformed metals visual inspection of the GND density map (Fig 11(a)) suggests that higher values are often located close to GBs. This was confirmed statistically by correlating the distance to GB (Fig. 8(d)) pixel by pixel with the measured GND density (Fig 11(c)). Despite significant scatter there is a clear trend of greater GND density closer to the GBs, suggesting that plastic deformation is occurred near the GB between the transformed martensite and the neighbouring grain (either austenite or martensite). Fig 11(d) shows how the average GND density within a grain is found to vary with grain size and although a larger number of data points would be desirable, the grains with high GND density ($> 4\times10^{15}\ m^{-2}$) are quite small in size. Morsdorf *et al* reported that smaller laths contain slightly higher density of dislocations than larger laths in Fe-0.13C-5.1Ni lath martensite [23]. This is because as martensitic transformation proceeds, the dislocation density increases in the remaining austenite and some of this dislocation is inherited to the martensite lath transformed from the dislocated austenite region [41, 42].

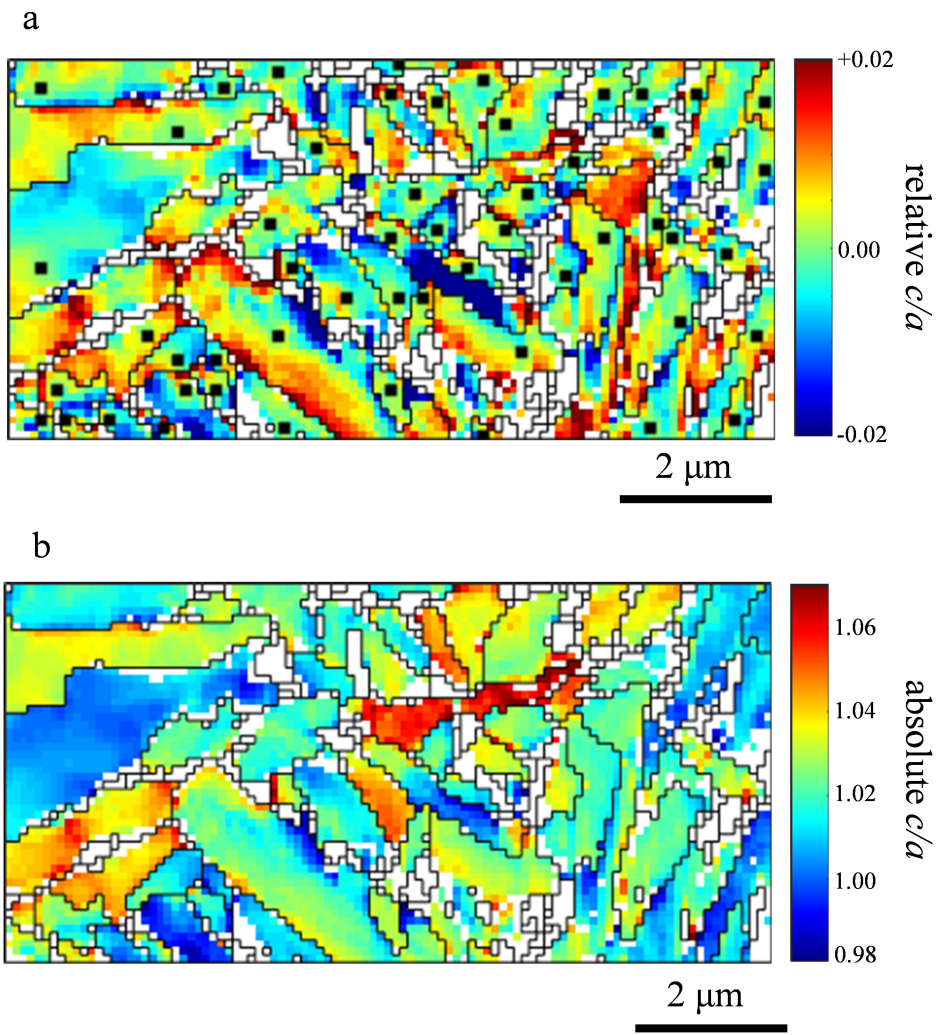

**Figure 9**. (a) Relative and (b) absolute *c/a* ratio map. The reference points are marked as the black squares in (a).

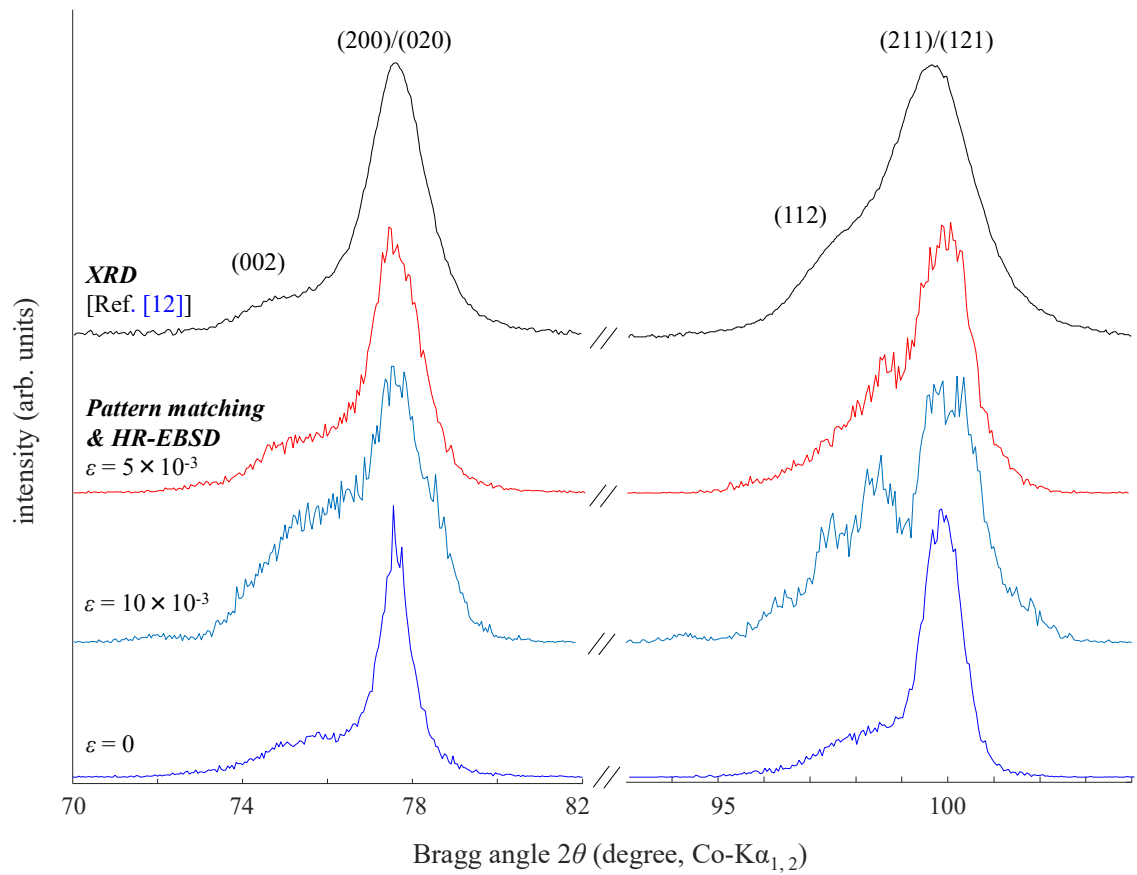

**Figure 10.** Lattice parameter frequency profile calculated on the assumption that the average and lattice strain of *a, b*-axes lengths at all reference points are 0.2856 *nm* [12] and 0, 5×10$^{-3}$, 10×10$^{-3}$ respectively. The lattice parameter is converted to the Bragg angle 2$\theta$. For comparison, the XRD profile from the same material [12] is also shown.

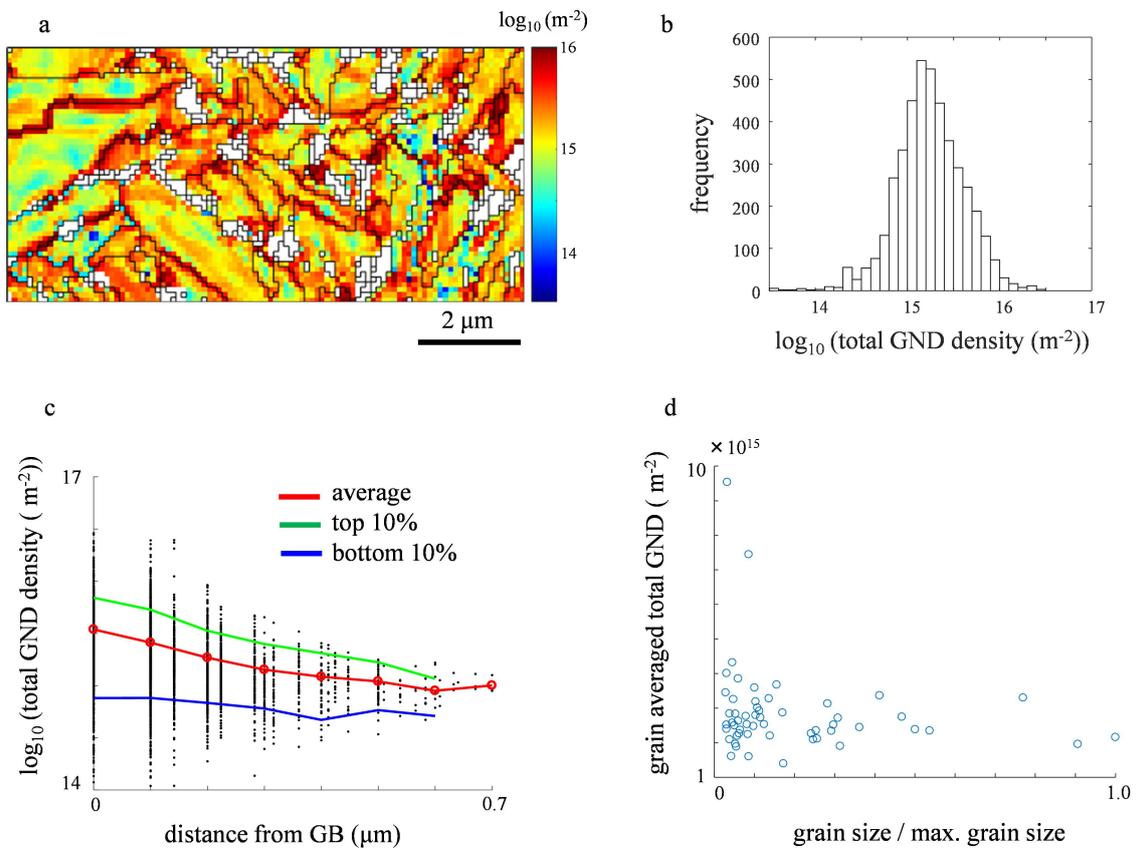

Figure 11. (a) Total GND density map and (b) histogram of the GND density frequency. GND density as a function of (c) the shortest distance to GBs and (d) grain size divided by the maximum of grain size in the map.

Table 5. Averaged and standard deviation of relative/absolute *c/a* ratio and GND density.

|  | average | ± 1 standard deviation |
|---|---|---|
| relative *c/a* | 0.000 | 0.009 |
| absolute *c/a* | 1.021 | 0.015 |
| total GND density (m$^{-2}$) | 2.09 ×10$^{15}$ | 2.43 ×10$^{15}$ |

It is likely that tetragonality (*c/a*), GND density and spatial location within the microstructure are all correlated. Fig. 12(a) shows that the relative *c/a* ratio tends to deviate most strongly from the grain average value (either increased or decreased) within a grain in regions where the GND density is larger. Since we have already shown that the GND density tends to be higher near GBs (Fig. 11(c)), it is not surprising that the large deviations of *c/a* from the grain average are also correlated with proximity to GBs (Fig. 12(b)). Interestingly, however, the absolute *c/a* ratio does not seem to correlate with the GND density (Fig. 12(c)). Since the local *c/a* ratio can be either decreased or increased from the grain average value by dislocations, the effect of dislocations on the absolute *c/a* ratio is balanced out overall. The absolute *c/a* ratio is also independent of the distance to GBs when the distance is smaller than 0.3 μm, although a slight decrease in the *c/a* is observed when the distance > 0.4 μm (Fig. 12(d)). This is because the *c/a* ratio tends to be decreased as the grain size increases. Fig. 13 shows grain averaged absolute *c/a* ratio with respect to the grain size. The grains with averaged *c/a* ratio > 1.04 are small in size, which is in good agreement with Fig. 12(d). However, some grains with the absolute *c/a* < 1 are also small in size. This might be related to the problem with grain size assessment in a two dimensional map, especially for small sized grains. A three dimensional observation would give more accurate picture for the grain size – *c/a* ratio relationship.

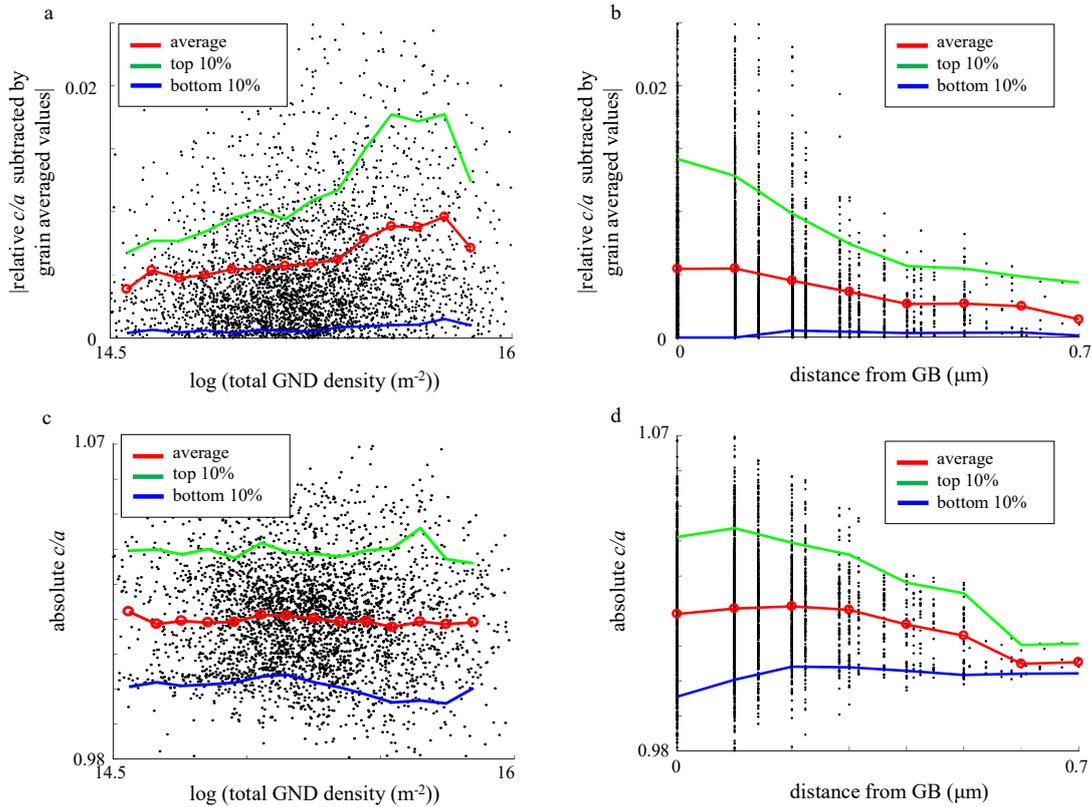

**Figure 12**. Scatter plot of relative *c/a* minus grain averaged *c/a* as functions of (a) total GND density and (b) the shortest distance to the nearest GBs. Absolute *c/a* at each pixel is plotted against (c) total GND density and (d) the shortest distance to the nearest GBs.

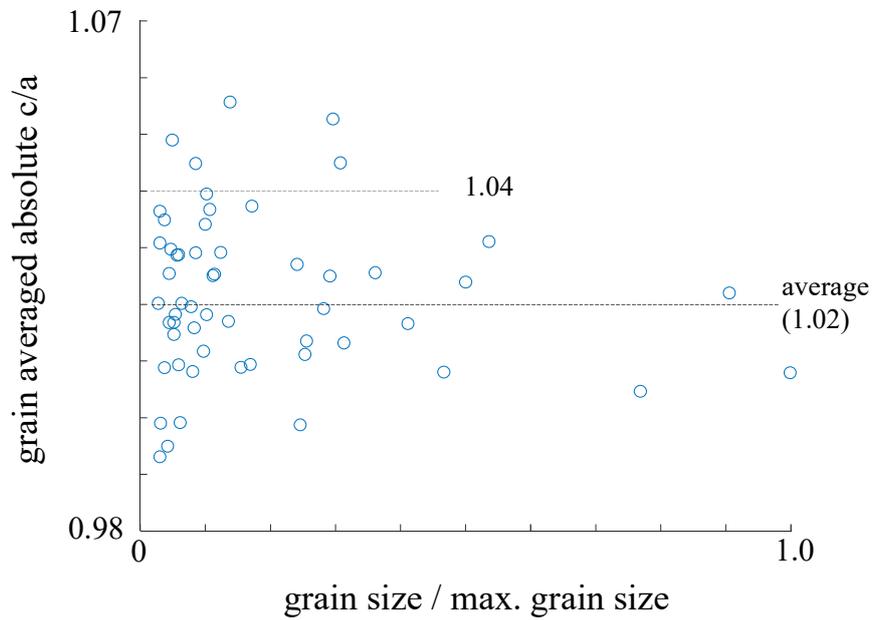

**Figure 13**. Grain averaged absolute *c/a* with respect to grain size.

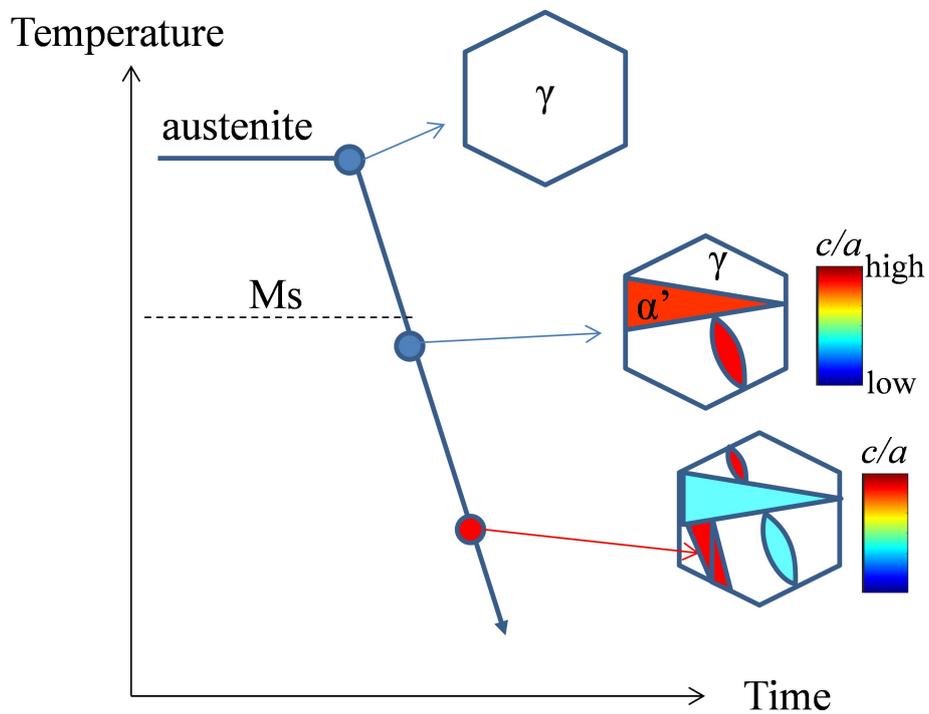

**Figure 14**. Schematic diagram of the formation of *c/a* variation in quenched Fe-C martensite during cooling.

## 3-4. Closing Discussion

We rationalise the tetragonality (*c/a*) and GND density distributions results reported here, along with other observations in the literature though the following processes occurring during cooling from the austenitising temperature (Fig. 14). In the austenite phase field carbon atoms sit at the centre of octahedral interstices as the volume of these sites is bigger than tetrahedral interstices. On cooling through the *Ms* temperature the Bain correspondence or Kurdjumov-Sacks (K-S) orientation relationship [43], results in the carbon atoms occupy octahedral *z* ($O_z$) sites in martensite [44] immediately after transformation. At this the moment when the martensite is first formed it thus must have the maximum tetragonal distortion. This was recently observed by in-situ neutron diffraction measurement for Fe-0.4C steel [34].

As the cooling continues toward room temperature, carbon atoms can diffuse within the martensite and form clusters and precipitates, and/or can segregate to GBs and dislocations owing to the auto-tempering [22, 23]. Furthermore the atomic positions of some of the interstitial carbon can be altered from $O_z$ site to other octahedral sites ($O_x$ and $O_y$ sites) and even tetrahedral sites to some extent, which can be promoted by the presence of compressive stresses [37]. This leads to a decrease in solid solution carbon and carbon order parameter [35-37], which contributes to both a decrease in the *c/a* ratio and to a reduction of the crystal symmetry. Pattern matching of EBSPs employed in this study assumed that the crystal structure of Fe-C martensite is either BCC or BCT (with varying *c/a*). Consideration of orthorhombic or lower symmetry structures might give better correlations between experimental and simulated patterns. A decrease in the length of *c*-axis upon quenching can lead to a flip in the longest axis among *a*-, *b*-, *c*-axes because of large elastic strains.

The exposure to time at temperature is reduced for martensite blocks that form later during cooling or sub-zero treatment and for these the extent of the auto-tempering effect should be smaller [23] and thus the decrease in the *c/a* ratio is relatively suppressed. The martensite blocks formed at the later stage of cooling also tend to be smaller in size [23, 45]. This explains our observation that the smaller grains (blocks) tend to show larger *c/a* ratio (Figs. 9(b) and 13). It is also reported that the *c/a* ratio of the martensite formed upon sub-zero treatment is larger than that of the martensite formed upon quenching to RT [46]. For those martensite blocks a flip in the longest axis is unlikely to occur.

The spatial variation of the *c/a* ratio was also observed within blocks, and this correlated with the GND density and proximity to GBs. Heterogeneous distribution of mechanically driven residual strains also contributes to the variation in the relative *c/a* ratio within a grain. However, the variation within a grain tends to be smaller than the one found between grains (blocks). A decrease in solid solution carbon content by the auto-tempering [22, 23] is thought to have a stronger influence than residual strains in causing the decrease in the *c/a* ratio. This variation in tetragonality and reduction in the crystal symmetry in Fe-C martensite is the origin of the complex X-ray diffraction profile from the martensite [12, 20, 21, 47]. The BCT Fe {200} XRD peak has two main sub-peaks originating from (200)/(020) and from (002) when carbon content is beyond the threshold value of ~0.4 wt.% for the onset of tetragonality. However, additional peak(s) in between the two main peaks should be taken into account to fully explain the whole XRD profile [12, 20, 21, 47]. This corresponds to the presence of lower *c/a* region or local distorted region in the Fe-C martensite, which is experimentally observed by the combination of pattern matching of EBSPs and HR-EBSD (Figs. 9 and 10). This raises a concern about the recent application of the Rietvelt refinement to the XRD profile analysis of Fe-C martensite using a simple BCT model [48-50]. Varying tetragonality including BCC and type II&III residual strains need to be taken into account for the more complete fitting, especially when analysisng low carbon martensite.

It is noted that the schematic of *c/a* distribution in Fig. 14 is expected to be different depending on carbon contents. When the carbon content is high (say more than 1 %) where the *Ms* temperature come close to RT, the extent of auto-tempering becomes smaller and uniform. Then the tetragonality is expected to be uniform. This can be seen in the XRD {200} profile of high carbon martensite, where (002)/(200) peak intensity ratio come close to the ideal value of 0.5 [51]. However the intensity ratio is known to decrease upon RT aging time for high carbon martensite [21, 51].

On the contrary, when the carbon content is lower, *c/a* ratio heterogeneities would be enhanced because of the higher *Ms* temperature and the crystal structure in an XRD analysis volume can be a mixture of BCC and BCT with varying tetragonality. The entire region can be regarded as BCC when the carbon content is below the threshold value, which was ~0.4 wt.% C for our quenching condition [12] and this should be dependent on the cooling rate [52].

The heterogeneities of crystal structure and distortion (*i.e.* residual strains) within a block/between blocks are considered to affect the mechanical properties of Fe-C marnteisitic steels, such as an initial yielding behaviour [8, 9]. Further investigation on this topic using quantitative HR-EBSD will be performed to better understand and control of the microstructure and mechanical properties of Fe-C martensitic steels.

## 4. Conclusions

In this study the distribution of *c/a* ratio in Fe-0.77C quenched martensite was investigated by the combination of pattern matching of EBSPs and cross-correlation based HR-EBSD. The *c/a* ratio analysis was also attempted using conventional Hough transform based analysis. The following findings are obtained:

(1) A Hough transform based conventional EBSD indexing was not able to distinguish the *c*-axis from the *a, b*-axes reliably in Fe-0.77C quenched martensite.

(2) Pattern matching of EBSPs improved the success rate of *c*-axis assignment as the longest axes, although the success rate did not reach a 100% match with analysis based on the Bain correspondence. This is mainly because there are some grains in which the *c/a* ratio was close to unity at the selected reference point.

(3) Within a grain the *c/a* ratio tended to deviate most significantly from the grain average for regions in the vicinity of GBs. The GND density was also increased in the vicinity of GBs.

(4) The spatial variation in the *c/a* ratio between martensite blocks is more evident than that within a block. The *c/a* ratios in small blocks tend to be higher than that in larger blocks. This suggests that the small blocks form later and have less redistribution of carbon atoms during the auto-tempering process so that the local *c/a* ratio remains high and crystal symmetry stays close to tetragonal.

(5) The variations in (i) block size, (ii) the tetragonality driven by local carbon content and ordering, (iii) the residual mechanical lattice strain, and (iv) the GND density all contribute to the variation in strength giving the martensite a relatively low stress for incipient plasticity, followed by rapid hardening to high proof, and ultimate tensile strengths.

## Acknowledgements

The authors express sincere thanks to Mr. Michiaki Matsumoto (Mishima Kosan Corporation) for his help with the SEM sample preparation. Prof. B. Hutchinson (Swerim AB) is also acknowledged for fruitful discussions about the residual stresses and tetragonality in Fe-C martensite. The research made use of facilities with the David Cockayne Centre for Electron Microscopy, Department of Materials, University of Oxford.